\documentclass[a4paper]{article}

\usepackage{INTERSPEECH2021}
\usepackage{adjustbox}
\usepackage{multirow}
\usepackage{multicol}
\usepackage{amssymb}
\usepackage{subcaption}
\usepackage{url}
\newlength{\bibitemsep}
\newlength{\bibparskip}
\let\oldthebibliography\thebibliography
\renewcommand\thebibliography[1]{%
  \oldthebibliography{#1}%
  \setlength{\parskip}{\bibitemsep}%
  \setlength{\itemsep}{\bibparskip}%
  \small
}

\title{Generalized Dilated CNN Models for Depression Detection Using Inverted Vocal Tract Variables}
\name{Nadee Seneviratne$^1$, Carol Espy-Wilson$^1$}
\address{
  $^1$University of Maryland - College Park}
\email{nadee@umd.edu, espy@umd.edu}

\begin{document}

\maketitle
\begin{abstract}
Depression detection using vocal biomarkers is a highly researched area. Articulatory coordination features (ACFs) are developed based on the changes in neuromotor coordination due to psychomotor slowing, a key feature of Major Depressive Disorder. However findings of existing studies are mostly validated on a single database which limits the generalizability of results. Variability across different depression databases adversely affects the results in cross corpus evaluations (CCEs). We propose to develop a generalized classifier for depression detection using a dilated Convolutional Neural Network which is trained on ACFs extracted from two depression databases. We show that ACFs derived from Vocal Tract Variables (TVs) show promise as a robust set of features for depression detection. Our model achieves relative accuracy improvements of $\sim10\%$ compared to CCEs performed on models trained on a single database. We extend the study to show that fusing TVs and Mel-Frequency Cepstral Coefficients can further improve the performance of this classifier.
\end{abstract}

\section{Introduction}
Major Depressive Disorder (MDD) is a mental health disorder that is characterized by long-lasting depressed mood or loss of interest in activities that will cause significant impairment in daily life. There are about 264 million people worldwide who suffer from depression \cite{WHO2020}. The serious consequences of MDD such as suicidality necessitates the need of reliable automated solutions that could help clinicians and therapists diagnose and treat MDD patients early and effectively and help patients in monitoring themselves. Previous studies have shown that vocal biomarkers developed using prosodic, source and spectral features \cite{CUMMINS201510,Scherer2013,Cummins2013b} can be very useful in depression detection. There are multiple studies that have performed the detection of depression using various combinations of speech features such as Mel Frequency Cepstral Coefficients (MFCCs), formants and voice quality features \cite{Cummins2011, Jiang2018}. Prior to developing DNN models based on network architectures such as Convolutional Neural Networks (CNNs) and Long Short Term Memory (LSTM) networks, early studies developed machine learning models mostly based on Support Vector Machines (SVM) and Gaussian Mixture Models (GMM).

Several recent studies found that successful results can be achieved by quantifying the changes in articulatory coordination to distinguish depressed speech from non-depressed speech \cite{Williamson2014,WILLIAMSON2019, Espy-Wilson2019, Seneviratne2020}. This difference in the timing of speech gestures is caused by a neurological phenomenon called psychomotor slowing, which is identified as a major characteristic of depression \cite{Whitwell1937}. It is viewed as a necessary feature of MDD and a key component in evaluating severity of depression \cite{ManualMentalDisOrd, WIDLOCHER198327}. Changes caused in speech due to psychomotor slowing such as more and longer pauses, slowed responses and monotonic phrases \cite{Sobin1997} lead to the usage of Articulatory Coordination Features (ACFs) to evaluate the severity of depression. ACFs are found to effectively capture information that can distinguish depressed speech from non-depressed speech using the multi-scale structure of correlations among the time series signals. This approach was predominantly validated using acoustic features such as formants and MFCCs as a proxy for underlying articulatory coordination \cite{WILLIAMSON2019, Williamson2016}. In our previous studies \cite{Espy-Wilson2019, Seneviratne2020} we showed that ACFs derived from direct articulatory speech features known as Vocal Tract Variables (TVs) are more effective in classifying depressed speech from non-depressed speech. These studies used the eigenspectra derived from the time-delay embedded correlation matrices as ACFs. An SVM classifier was used due to the limited availability of data. The magnitudes of the eigenvalues of the eigenspectra derived from the time-delay embedded correlation matrix showed the complexity of articulatory coordination. The study in \cite{Huang2020} explains that this channel-delay correlation matrix can be further optimized to eliminate repetitive sampling and matrix discontinuities. A more effective and scalable representation for ACFs was proposed in \cite{Huang2020} utilizing dilated CNNs and more delays in the correlation matrix. 

With the advent of Deep Neural Networks (DNN), its applications in speech based depression detection and severity prediction increased rapidly, yielding promising results \cite{Yang2017, Yin2019, Ray2019}. However, the generalizability of these models is limited provided that these studies were performed on a single database. The characteristics of available depression databases differ depending on the acoustic variability in the speech recordings due to different speech types (free, read, sustained vowels etc.), speaking styles and rates, speaker demographics and different types of studies (observational studies/clinical trials) etc. Thus, findings from one study may not always be observed across different databases even within the same language. Therefore, the need to develop more generalized models prevails. Domain  adaptation techniques have been explored to address this issue \cite{Huang2020IS}.

In this paper, we present a novel approach of using these 'direct' articulatory parameters (TVs) in a deep learning setting for the first time to detect depression. We combine speech data from two depression databases with different characteristics to develop generalized CNN models to detect the presence of depression. Using the approach proposed in \cite{Huang2020}, we show that robust and generalized Dilated CNN models can be developed using the TV based ACFs to perform this task.

\section{Database Descriptions}
% \vspace{-5pt}
Speech data from two databases were combined for our experiments. We encountered two clinician (CL)-rated depression assessment scales used in these databases: Hamilton Depression Rating Scale (HAMD) and Quick Inventory of Depressive Symptomatology (QIDS). The severity level definition for each class can be found in Table \ref{tab:dep_severity}. Data in levels 2-5 is combined for the `depressed' category and data in level 1 is used for the `non-depressed' category. 

\begin{table}[h]
    \centering
    \small
    \caption{\textit{Severity level definitions of MDD assessment scales}}
    \label{tab:dep_severity}
    \begin{adjustbox}{max width = \columnwidth}
        \begin{tabular}{ccc}
            \toprule
            \textbf{Severity Level} & \textbf{HAMD} & \textbf{QIDS} \\ \midrule
            1. Normal & 0 – 7 & 0 - 5 \\ 
            2. Mild & 8 - 13 & 6 - 10 \\ 
            3. Moderate & 14 - 18 & 11 - 15 \\
            4. Severe & 19 - 22 & 16 - 20 \\ 
            5. Very Severe & 23 - 52 & 21 - 27 \\ \bottomrule
        \end{tabular}
    \end{adjustbox}
    % \vspace{-10pt}
\end{table}

% \vspace{-10pt}

\begin{table}[h]
    \centering
    \caption{\textit{Details of Depression Databases}}
    \label{tab:dep_databases}
    \begin{adjustbox}{max width = \columnwidth}
        \begin{tabular}{ccc}
        \toprule
        \multicolumn{1}{c}{\textbf{Database}} & \multicolumn{1}{c}{\textbf{MD-1}} & \multicolumn{1}{c}{\textbf{MD-2}} \\ \midrule
        Longitudinal & 6 Weeks & 4 Weeks \\ 
        % Study Type & Observational & Clinical trial \\
        \# Subjects & 20 F, 15 M & 104 F, 61 M \\ 
        Demography & 31 Caucasian & 125 Caucasian \\
         & 1 African American & 26 African American \\
         & 1 Bi-racial & 4 Asian \\
         & 1 Greek, 1 Hispanic & 10 Other \\ 
        Assessment & HAMD-CL: Bi-weekly & HAMD-CL, QIDS-CL: Weeks 1,2,4 \\ 
        FS Lengths & \multicolumn{1}{c}{Min: 2.5s, Max: 156.8s} & \multicolumn{1}{c}{Min: 2.6s, Max: 181.2s} \\ 
        Recording Type & \multicolumn{2}{c}{Interactive Voice Response Technology (8kHz)} \\ \bottomrule
        \end{tabular}
    \end{adjustbox}
    % \vspace{5pt}
\end{table}

Details of the depression databases are given in Table \ref{tab:dep_databases}. In addition, MD-1 is an observational study where patients started on pharmacotherapy and/or psychotherapy treatment for depression close to the beginning of the study. MD-2 is a clinical trial where subjects weren't taking psychotropic medications at the beginning of the study and started on 50 mg/day of sertraline or placebo (double-blind and randomized) at baseline. In this study, we used recordings of free speech (FS) where patients describe how they feel emotionally, physically and their ability to function in each week. Additionally these databases contain read speech recordings which were not used for this study. In MD-2, depression assessment scores were provided to only 105 subjects. Due to the availability of two CL-rated scores in MD-2, only the speech samples where both the scores belong to the same category were used.

% \vspace{-10pt}
\section{Articulatory Coordination Features}
% \vspace{-5pt}
\subsection{Speech Inversion (SI) System}
% \vspace{-5pt}
In Articulatory Phonology (AP) \cite{Browman1992}, speech is viewed as a constellation of overlapping gestures. These gestures are discrete action units whose activation results in constriction formation or release by five distinct constrictors along the vocal tract: lips, tongue tip, tongue body, velum, and glottis). The vocal tract variables (TVs) are defined by the constriction degree and location of these five constrictors. A speaker independent, DNN based SI system is used to compute the trajectory of the TVs that represent constriction location and degree of articulators located along the vocal tract \cite{Sivaraman2019}. The model was trained using the Wisconsin X-Ray Microbeam (XRMB) database \cite{Westbury1994a}. The six TVs estimated by the SI system are – Lip Aperture, Lip Protrusion, Tongue Body Constriction Location, Tongue Body Constriction Degree, Tongue Tip Constriction Location and Tongue Tip Constriction Degree. For detailed information about the SI system, the reader is referred to \cite{Sivaraman2019}.

\emph{\textbf{Glottal TV Estimation: }}
For a complete representation of TVs described in the AP, TVs related to the glottal state need to be included. Due to the difficulty in placing sensors near the glottis to acquire ground-truth information, the DNN based system could not be trained using articulatory data.
Hence, we used the periodicity and aperiodicity measure obtained from the Aperiodicity, Periodicity and Pitch detector developed in \cite{Deshmukh2005}. This program estimates the proportion of periodic energy and aperiodic energy and pitch of a speech signal based on the distribution of the minima of the average magnitude difference function of the speech signal. In \cite{Seneviratne2020}, these glottal parameters boosted the classification accuracies for depressed and non-depressed speech classification by about 8\%.

\subsection{MFCCs Estimation}
% \vspace{-5pt}
We use ACFs derived from MFCCs as a proxy for actual articulatory features to enable fair comparisons with the experiments conducted using TV based ACFs. For this, 12 MFCC time series were extracted by using an analysis window of 20 ms with a 10 ms frame shift (1\textsuperscript{st} MFCC coefficient was discarded). 

% \vspace{-10pt}
\subsection{Channel-delay Correlation Matrix}
% \vspace{-5pt}
Conventionally for an $M$-channel feature vector (TVs or MFCCs), the channel-delay correlation matrix in \cite{Seneviratne2020} has dimensionality of ($kM \times kM$) for $k (15)$ time delays per channel with a fixed delay scale (7 samples). In order to incorporate multiple delay scales $p$, these matrices computed at different delay scales will be stacked yielding a $p \times kM \times kM$ dimensional matrix \cite{WILLIAMSON2019}. In \cite{Huang2020}, the authors propose a novel method to construct the channel-delay correlation matrix that overcomes the limitations found in the conventional approach such as repetitive sampling and matrix discontinuities at the borders of adjacent sub-matrices. %Hence, in this work, we have adopted this new structure of channel-delay correlation matrix which will make use of dilated CNN layers to incorporate multiple delay scales.

In this work, we have adopted this new structure of channel-delay correlation matrix.
For an $M$-channel feature vector $\mathbf{X}$, the delayed correlations ($r_{i,j}^d$) between $i^{th}$ channel $\mathbf{x_i}$ and $j^{th}$ channel $\mathbf{x_{j}}$ delayed by $d$ frames, are computed as:
% \vspace{-5pt}
\begin{equation}
    r_{i,j}^d = \frac{\sum_{t=0}^{N-d-1}x_i[t]x_j[t+d]}{N-|d|}
\end{equation}
where N is the length of the channels.
The correlation vector for each pair of channels with delays $d \in [0,D]$ frames will be constructed as follows:
% \vspace{-5pt}
\begin{equation}
    R_{i,j} = \begin{bmatrix}r_{i,j}^0, & r_{i,j}^1, & \dots & r_{i,j}^D \end{bmatrix}^T \in \mathbb{R}^{1\times (D+1)}
\end{equation}
%\vspace{-5pt}
The delayed auto-correlations and cross-correlations are stacked to construct the channel-delay correlation matrix:
\begin{equation}
    \mathit{\widetilde{R}_{ACF}} =  \begin{bmatrix}R_{1,1} & \dots & R_{i,j} & \dots & R_{M,M}\end{bmatrix}^T \in \mathbb{R}^{M^2\times (D+1)}
\end{equation}

It is important to note that the $\mathit{\widetilde{R}_{ACF}}$ matrix contains every correlation only once. With this representation, information pertaining to multiple delay scales can be incorporated into the model by using dilated CNN layers with corresponding dilation factors while maintaining a low input dimensionality. Each $R_{i,j}$ will be processed as a separate input channel in the CNN model, and thereby overcoming discontinuities.

\section{Dilated CNN Classifier}
\vspace{-7pt}
 
\begin{figure*}[t]
    \centering
    \includegraphics[width=\textwidth]{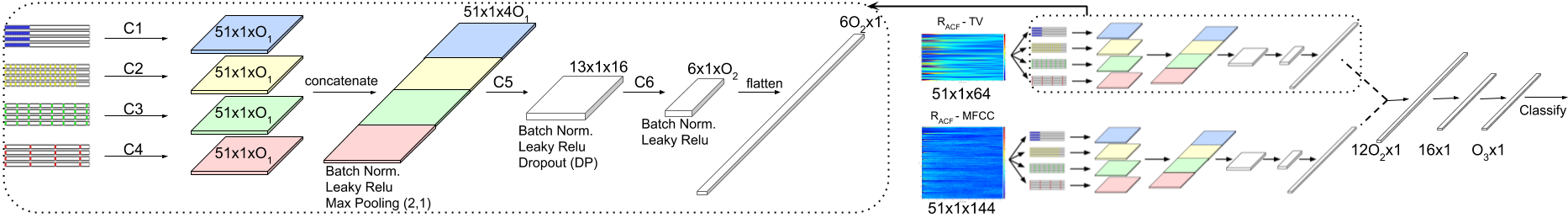}
    \caption{\textit{Dilated CNN Architecture for TV and MFCC Fused Model}}
    \label{fig:dnn_model}
    \vspace{-15pt}
\end{figure*}

The Dilated CNN architecture proposed in \cite{Huang2020} was adopted in the experiments (Fig. \ref{fig:dnn_model}). The input $\mathit{\widetilde{R}_{ACF}}$ is fed into four parallel convolutional layers ($C1,C2,C3,C4$) with different dilation rates $n=\{1,3,7,15\}$ and a kernel size of $(15,1)$ which resembles the multiple delay scales in the conventional approach. The outputs of these four parallel layers are concatenated and then passed through two sequential convolutional layers $(C5,C6)$. This output is flattened and passed through two fully connected (dense) layers $(D1,D2)$ to perform 2-class classification in the output layer. All convolutional layers used LeakyRelu activation, whereas dense layers used Relu activation with $l_2$ regularization of $\lambda=0.01$. Batch Normalization, dropouts, and max-pooling layers were added as shown in the Fig. \ref{fig:dnn_model}. The weight sharing nature of CNNs handles the high dimensional correlation matrices with a low number trainable parameters.

% \vspace{-10pt}
\section{Experiments and Results}
% \vspace{-5pt}
\subsection{Data Preparation}
% \vspace{-5pt}
We conducted our experiments using FS since previous work \cite{Seneviratne2020} has shown a larger difference of articulatory coordination between depressed and non-depressed speech for FS than for read speech. This difference is probably due to the increased cognitive load associated with FS. To increase the number of samples to train the CNN model, we segmented the audio recordings that are longer than 20s into segments of 20s with a shift of 5s (to also make the CNN models resilient to translation). Recordings with duration less than 10s were discarded and other shorter recordings (between 10s-20s) were used as they were. Table \ref{tab:durations} summarizes the total duration of speech data available after the segmentation. The speech samples were split into train/validation/test sets ($80:10:10$) preserving a similar distribution between the 2 classes and ensuring that there is no-overlap in speakers in the train/validation/test sets. The class imbalance issue is addressed during the training process by assigning class-weights to both training and validation sets.
% \vspace{-10pt}
\begin{table}[h]
    \centering
    \small
    \caption{\textit{Duration of Available Data in hours}}
    \label{tab:durations}
    \begin{adjustbox}{max width = \columnwidth}
    \begin{tabular}{ccc}
        \toprule
        \textbf{Database} & \textbf{Depressed} & \textbf{Non-depressed} \\ \midrule
        MD-1 & 11.78 & 2.45 \\ 
        MD-2 & 15 & 1.2 \\ 
        MD-1\&2 & 26.78 & 3.65 \\ \bottomrule
        \end{tabular}
    \end{adjustbox}
\end{table}
\vspace{-10pt}
%% include speaker distribution????

CNN models were trained using two sets of features: TVs corresponding to constriction degree and location and glottal parameters - 8 channels and MFCCs - 12 channels. Before computing the correlation matrices, feature vectors were mean-variance normalized individually.

% \vspace{-10pt}
\subsection{Model Training}
% \vspace{-5pt}
All models were trained on $D=50$ which was empirically determined and a learning rate of $1e-5$. The models were optimized using the Adam Optimizer. Binary Cross Entropy was used as the loss function. For $C5$, kernel size was $(3,1)$ with a stride of $2$ and 16 filter outputs were used. For $C1-C5$, `same' padding was used and for $C6$ `valid' padding was used. All inputs were z-normalized using the mean and the standard deviation of the training data. All models were trained with early stopping criteria based on validation loss (patience 15 epochs) for a maximum of 300 epochs. To evaluate the performance of models, overall accuracy, area under the receiver operating characteristics curve (AUC-ROC), and F1 scores were used.

\begin{table}[h]
    \centering
    \small
    \caption{\textit{Grid Search Parameters - Best Model (for MD1\&2)}}
    \label{tab:hyperparam}
    \begin{adjustbox}{max width = \columnwidth}
    \begin{tabular}{lccccc}
    \toprule
     & \textbf{C1-C4 Filter} & \textbf{C6 Filter} & \textbf{C6 Kernel} & \textbf{D2 Output} & \textbf{Dropout} \\
     & \textbf{Outputs ($O_1$)} & \textbf{Output ($O_2$)} & \textbf{($K_1$)} & \textbf{Size ($O_3$)} & \textbf{Prob. ($DP$)} \\ \midrule
    \textbf{Range} & \{16,32\} & \{8,16\} & \{(3,1), (4,1)\} & \{8,16\} & \{0.4,0.5\} \\ 
    \textbf{TV} & 32 & 16 & (3,1) & 8 & 0.5 \\
    \textbf{MFCC} & 16 & 8 & (3,1) & 16 & 0.5 \\ 
    \textbf{Fused} & 32 & 8 & (3,1) & 8 & 0.5 \\ \bottomrule
    \end{tabular}
    \end{adjustbox}
    % \vspace{-13pt}
\end{table}

Grid search was performed to tune the hyper-parameters using the ranges in Table \ref{tab:hyperparam}. Models were trained with train-test combinations as shown in Table \ref{tab:2cls_results}. Note that MD-1\&2 represents the combination of MD-1 and MD-2 data.

% \vspace{-10pt}
\subsection{Classification Results}
% \vspace{-8pt}
\label{sec:class_res}
The results are included in Table \ref{tab:2cls_results}. We evaluate the performance of our models using AUC-ROC for fair comparison. Accuracy alone could not be used as an indication of model strength due to class imbalance. Compared to cross corpus evaluations (CCEs) of TV based models trained on a single database (i.e. train/test pairs of MD-2/MD-1 \& MD-1/MD-2), the AUC-ROCs of the combined model (MD-1\&2/MD-1 \& MD-1\&2/MD-2) indicate relative improvements of $\sim$19\% and $\sim$14\% respectively. For MFCC based models, the corresponding relative improvements are $\sim$12\% and $\sim$15\%. Therefore we believe the model trained on combined data performs better compared to single corpus models.
Further, MFCC based combined model seems to outperform MFCC based models that were trained and evaluated on the same database through learning on additional data. Due to generalization it is possible that with the combined data model performance for a particular database be sometimes slightly lower compared to the models trained solely on the same database as can be seen in TV based models.

% \vspace{-15pt}
\begin{table}[h]
\caption{\textit{Results for 2-Class Classifications}}
\label{tab:2cls_results}
\renewcommand{\arraystretch}{1.25}
\begin{adjustbox}{max width = \columnwidth}
\begin{tabular}{cccccc}
\toprule
\textbf{Feats} & \textbf{Train} & \textbf{Test} & \textbf{Accuracy} & \textbf{\small{AUC-ROC}} & \textbf{F1(D)/F1(ND)} \\ \midrule
\multirow{7}{*}{TV} & MD-1\&2 & MD-1\&2 & 82.67\% & 0.7329 & 0.9/0.42 \\
 & MD-1\&2 & MD-1 & 78.46\% & 0.7298 & 0.87/0.44 \\  
 & MD-1\&2 & MD-2 & 86.39\% & 0.7309 & 0.92/0.39 \\ 
 & MD-1 & MD-1 & 82.31\% & 0.7757 & 0.9/0.41 \\ 
 & MD-1 & MD-2 & 78.57\% & 0.6399 & 0.87/0.26 \\
 & MD-2 & MD-2 & 85.71\% & 0.8014 & 0.92/0.43 \\
 & MD-2 & MD-1 & 71.92\% & 0.6127 & 0.82/0.37 \\ \hline
\multirow{7}{*}{MFCC} & MD-1\&2 & MD-1\&2 & 81.77\% & 0.7447 & 0.89/0.36 \\ 
 & MD-1\&2 & MD-1 & 74.62\% & 0.6249 & 0.85/0.3 \\  
 & MD-1\&2 & MD-2 & 88.10\% & 0.9141 & 0.93/0.44 \\ 
 & MD-1 & MD-1 & 76.54\% & 0.6190 & 0.86/0.33 \\
 & MD-1 & MD-2 & 81.29\% & 0.7923 & 0.89/0.25 \\
 & MD-2 & MD-2 & 80.61\% & 0.7353 & 0.89/0.28 \\
 & MD-2 & MD-1 & 78.46\% & 0.5583 & 0.88/0.18 \\ \bottomrule
\end{tabular}
\end{adjustbox}
% \vspace{-10pt}
\end{table}
% \vspace{-13pt}
\vspace{-10pt}
\subsection{Classification using Fused TVs and MFCCs}
% \vspace{-5pt}
A fused model was trained combining TVs and MFCCs in order to investigate the inter-learning among ACFs derived from different feature vectors. Adopting a late fusion strategy, the model consists of two parallel structures similar to the previous model up to $C6$ using ACFs derived from MFCCs and TVs as inputs. Late fusion strategy was adopted to avoid the input feature dimensions being unnecessary large which may lead to overfitting issues. 

The outputs of the $C6$ layers of both structures were flattened and concatenated and then passed through two dense layers to perform the classification. Similar to section \ref{sec:class_res}, the general trend of the combined model performing better than the CCEs of single corpus models is evident through the relative AUC-ROC improvements of $\sim$14\% and $\sim$18\% for MD-1 data and MD-2 data respectively (Table \ref{tab:fused_results}). Additionally, fused model generally shows better AUC-ROC values when trained on the combined data compared to TV based and MFCC based models.

\begin{figure}[t]
    \centering
    \begin{subfigure}{0.4\columnwidth}
    \includegraphics[width=\linewidth]{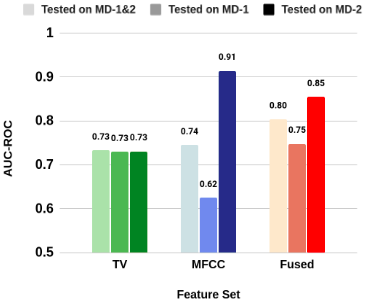}
    \subcaption{}
    \label{fig:md12-aucroc}
    \end{subfigure}
    \begin{subfigure}{0.25\columnwidth}
    \includegraphics[width=\linewidth]{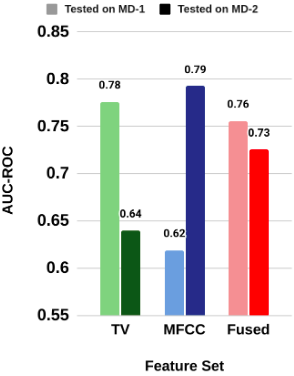}
    \subcaption{}
    \label{fig:md1-aucroc}
    \end{subfigure}
    \begin{subfigure}{0.25\columnwidth}
    \includegraphics[width=\linewidth]{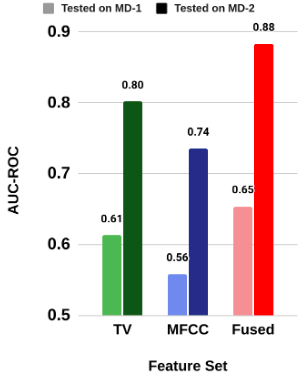}0
    \subcaption{}
    \label{fig:md2-aucroc}
    \end{subfigure}
    \caption{AUC-ROC for Models Trained on (a) MD-1\&2, (b) MD-1 only and (c) MD-2 only when evaluated on MD-1\&2 (for (a) only), MD-1 and MD-2 for 3 Feature Sets.}
    \label{fig:auc-roc}
\end{figure}

% \vspace{-10pt}
\begin{table}[h]
\caption{\textit{Results for Fused Models Classifications}}
\label{tab:fused_results}
\renewcommand{\arraystretch}{1.25}
\centering
\small
\begin{adjustbox}{max width = \columnwidth}
\begin{tabular}{ccccc}
\toprule
\textbf{Train} & \textbf{Test} & \textbf{Accuracy} & \textbf{AUC-ROC} & \textbf{F1(D)/F1(ND)} \\ \midrule
MD-1\&2 & MD-1\&2 & 87.73\% & 0.8036 & 0.93/0.39 \\ 
MD-1\&2 & MD-1 & 83.07\% & 0.7479 & 0.9/0.39 \\
MD-1\&2 & MD-2 & 91.84\% & 0.8546 & 0.96/0.4 \\ 
MD-1 & MD-1 & 83.84\% & 0.7555 & 0.9/0.48 \\ 
MD-1 & MD-2 & 82.65\% & 0.7254 & 0.9/0.11 \\ 
MD-2 & MD-2 & 77.21\% & 0.8824 & 0.86/0.36 \\
MD-2 & MD-1 & 71.54\% & 0.6535 & 0.82/0.31 \\ \bottomrule
\end{tabular}
\end{adjustbox}
% \vspace{-5pt}
\end{table}

%  \vspace*{-25pt}
\vspace*{1pt}
\section{Discussion}
% \vspace{-5pt}
In this paper we showed that Dilated CNN based models trained for depression detection can be generalized by combining data from multiple databases using a concise representation of channel-delay correlation matrices of feature vectors. Relatively high AUC-ROC variance of MFCC based models across different experiments  indicate that the performance of MFCC based models are database-dependent compared to TV based models. TV based models seem to consistently perform with lower AUC-ROC variance (Figure \ref{fig:md12-aucroc}{}). Therefore, ACFs derived from TVs are desirable in general compared to those from MFCCs, indicating that TV based ACFs could act as more robust features in the generalization process. This observation is further supported as TVs provide a direct articulatory representation. Further fused model experiment suggests that combining both TVs and MFCCs helps to boost the classification performance of the generalized model.  

We observe that TV based models perform better on MD-1 data while MFCC based models perform better on MD-2 data. Fused models seem to capture the strengths of both TVs and MFCCs and thereby perform better on both MD-1 and MD-2 data. This result suggests that ACFs derived from different features complement each other.

The findings of the cross corpus studies are encouraging. The TV based models perform significantly better than the MFCC based models when trained and evaluated on the same database (Figures \ref{fig:md1-aucroc} and \ref{fig:md2-aucroc}). On the other hand, MFCC based models show higher accuracy gains when evaluated across databases. However, we note that the minority class (non-depressed) F1 scores are lower, suggesting that the model is biased towards the majority class given the class imbalance in test data. In comparison TV based models have higher F1 scores for the minority class suggesting that they generalize better on unseen data. This also leads to the hypothesis that the normalized TVs can provide a more speaker-independent representation compared to the MFCCs that results in TV based articulatory coordination features to provide stable results across the two databases compared to MFCCs.

In the future, this study will be extended to more databases. In the case of different MDD assessment scales, consideration needs to be paid to achieve uniformity in establishing ground-truth. In-depth analysis will be conducted to understand the differences observed for TV based ACFs and MFCC based ACFs.

% \vspace{-15pt}
% \vspace*{-10pt}
\section{Acknowledgements}
\vspace{-5pt}
We would like to thank Dr. James Mundt for the depression databases MD-1\&2 \cite{MUNDT2007, MUNDT2012}. We also thank Dr. Thomas Quatieri and Dr. James Williamson for granting access to the MD-2 database \cite{MUNDT2012} which was funded by Pfizer. This work was supported by the UMCP \& UMB Artificial Intelligence + Medicine for High Impact Challenge Award and a Faculty Student Research Award from UMCP.

\bibliographystyle{IEEEtran}
\bibliography{references}

\begin{thebibliography}{10}
\providecommand{\url}[1]{#1}
\csname url@samestyle\endcsname
\providecommand{\newblock}{\relax}
\providecommand{\bibinfo}[2]{#2}
\providecommand{\BIBentrySTDinterwordspacing}{\spaceskip=0pt\relax}
\providecommand{\BIBentryALTinterwordstretchfactor}{4}
\providecommand{\BIBentryALTinterwordspacing}{\spaceskip=\fontdimen2\font plus
\BIBentryALTinterwordstretchfactor\fontdimen3\font minus
  \fontdimen4\font\relax}
\providecommand{\BIBforeignlanguage}[2]{{%
\expandafter\ifx\csname l@#1\endcsname\relax
\typeout{** WARNING: IEEEtran.bst: No hyphenation pattern has been}%
\typeout{** loaded for the language `#1'. Using the pattern for}%
\typeout{** the default language instead.}%
\else
\language=\csname l@#1\endcsname
\fi
#2}}
\providecommand{\BIBdecl}{\relax}
\BIBdecl

\bibitem{WHO2020}
\BIBentryALTinterwordspacing
{World Health Organization (WHO)}, ``Depression,'' 2020. [Online]. Available:
  \url{https://www.who.int/news-room/fact-sheets/detail/depression}
\BIBentrySTDinterwordspacing

\bibitem{CUMMINS201510}
\BIBentryALTinterwordspacing
N.~Cummins, S.~Scherer, J.~Krajewski, S.~Schnieder, J.~Epps, and T.~F.
  Quatieri, ``A review of depression and suicide risk assessment using speech
  analysis,'' \emph{Speech Communication}, vol.~71, pp. 10 -- 49, 2015.
  [Online]. Available:
  \url{http://www.sciencedirect.com/science/article/pii/S0167639315000369}
\BIBentrySTDinterwordspacing

\bibitem{Scherer2013}
S.~{Scherer}, G.~{Stratou}, M.~{Mahmoud}, J.~{Boberg}, J.~{Gratch}, A.~{Rizzo},
  and L.~{Morency}, ``Automatic behavior descriptors for psychological disorder
  analysis,'' in \emph{2013 10th IEEE International Conference and Workshops on
  Automatic Face and Gesture Recognition (FG)}, 4 2013, pp. 1--8.

\bibitem{Cummins2013b}
N.~Cummins, J.~Epps, V.~Sethu, M.~Breakspear, and R.~Goecke, ``Modeling
  spectral variability for the classification of depressed speech,''
  \emph{Proceedings of the Annual Conference of the International Speech
  Communication Association, INTERSPEECH}, pp. 857--861, 01 2013.

\bibitem{Cummins2011}
N.~Cummins, J.~Epps, M.~Breakspear, and R.~Goecke, ``An investigation of
  depressed speech detection: Features and normalization.'' \emph{Proceedings
  of the Annual Conference of the International Speech Communication
  Association, INTERSPEECH}, pp. 2997--3000, 01 2011.

\bibitem{Jiang2018}
H.~Jiang, B.~Hu, Z.~Liu, G.~Wang, L.~Zhang, X.~Li, and H.~Kang, ``Detecting
  depression using an ensemble logistic regression model based on multiple
  speech features,'' \emph{Computational and Mathematical Methods in Medicine},
  vol. 2018, pp. 1--9, 09 2018.

\bibitem{Williamson2014}
\BIBentryALTinterwordspacing
J.~R. Williamson, T.~F. Quatieri, B.~S. Helfer, G.~Ciccarelli, and D.~D. Mehta,
  ``Vocal and facial biomarkers of depression based on motor incoordination and
  timing,'' in \emph{Proceedings of the 4th International Workshop on
  Audio/Visual Emotion Challenge}, ser. AVEC ’14.\hskip 1em plus 0.5em minus
  0.4em\relax New York, NY, USA: Association for Computing Machinery, 2014, p.
  65–72. [Online]. Available: \url{https://doi.org/10.1145/2661806.2661809}
\BIBentrySTDinterwordspacing

\bibitem{WILLIAMSON2019}
\BIBentryALTinterwordspacing
J.~R. Williamson, D.~Young, A.~A. Nierenberg, J.~Niemi, B.~S. Helfer, and T.~F.
  Quatieri, ``Tracking depression severity from audio and video based on speech
  articulatory coordination,'' \emph{Computer Speech \& Language}, vol.~55, pp.
  40 -- 56, 2019. [Online]. Available:
  \url{http://www.sciencedirect.com/science/article/pii/S0885230817303510}
\BIBentrySTDinterwordspacing

\bibitem{Espy-Wilson2019}
\BIBentryALTinterwordspacing
C.~Espy-Wilson, A.~C. Lammert, N.~Seneviratne, and T.~F. Quatieri, ``{Assessing
  Neuromotor Coordination in Depression Using Inverted Vocal Tract
  Variables},'' in \emph{Proc. Interspeech 2019}, 2019, pp. 1448--1452.
  [Online]. Available: \url{http://dx.doi.org/10.21437/Interspeech.2019-1815}
\BIBentrySTDinterwordspacing

\bibitem{Seneviratne2020}
N.~Seneviratne, J.~R. Williamson, A.~C. Lammert, and T.~F. Q.~C. Espy-Wilson,
  ``Extended study on the use of vocal tract variables to quantify neuromotor
  coordination in depression,'' in \emph{Proc. Interspeech 2020}, 2020.

\bibitem{Whitwell1937}
J.~R. Whitwell, \emph{Historical notes on psychiatry}.\hskip 1em plus 0.5em
  minus 0.4em\relax Oxford, England, 1937.

\bibitem{ManualMentalDisOrd}
\BIBentryALTinterwordspacing
{American Psychiatric Association}, \emph{Diagnostic and Statistical Manual of
  Mental Disorders}.\hskip 1em plus 0.5em minus 0.4em\relax Washington, DC,
  2000. [Online]. Available:
  \url{https://dsm.psychiatryonline.org/doi/abs/10.5555/appi.books.9780890425596.x00pre}
\BIBentrySTDinterwordspacing

\bibitem{WIDLOCHER198327}
\BIBentryALTinterwordspacing
D.~J. Widlöcher, ``Psychomotor retardation: Clinical, theoretical, and
  psychometric aspects,'' \emph{Psychiatric Clinics of North America}, vol.~6,
  no.~1, pp. 27 -- 40, 1983, recent Advances in the Diagnosis and Treatment of
  Affective Disorders. [Online]. Available:
  \url{http://www.sciencedirect.com/science/article/pii/S0193953X18308384}
\BIBentrySTDinterwordspacing

\bibitem{Sobin1997}
C.~Sobin and H.~Sackeim, ``Psychomotor symptoms of depression,'' \emph{The
  American journal of psychiatry}, vol. 154, pp. 4--17, 02 1997.

\bibitem{Williamson2016}
\BIBentryALTinterwordspacing
J.~R. Williamson, E.~Godoy, M.~Cha, A.~Schwarzentruber, P.~Khorrami, Y.~Gwon,
  H.-T. Kung, C.~Dagli, and T.~F. Quatieri, ``Detecting depression using vocal,
  facial and semantic communication cues,'' in \emph{Proceedings of the 6th
  International Workshop on Audio/Visual Emotion Challenge}, ser. AVEC
  ’16.\hskip 1em plus 0.5em minus 0.4em\relax New York, NY, USA: Association
  for Computing Machinery, 2016, p. 11–18. [Online]. Available:
  \url{https://doi.org/10.1145/2988257.2988263}
\BIBentrySTDinterwordspacing

\bibitem{Huang2020}
\BIBentryALTinterwordspacing
Z.~Huang, J.~Epps, and D.~Joachim, ``Exploiting vocal tract coordination using
  dilated {CNNS} for depression detection in naturalistic environments,'' in
  \emph{2020 {IEEE} International Conference on Acoustics, Speech and Signal
  Processing, {ICASSP} 2020, Barcelona, Spain, May 4-8, 2020}.\hskip 1em plus
  0.5em minus 0.4em\relax {IEEE}, 2020, pp. 6549--6553. [Online]. Available:
  \url{https://doi.org/10.1109/ICASSP40776.2020.9054323}
\BIBentrySTDinterwordspacing

\bibitem{Yang2017}
\BIBentryALTinterwordspacing
L.~Yang, H.~Sahli, X.~Xia, E.~Pei, M.~C. Oveneke, and D.~Jiang, ``Hybrid
  depression classification and estimation from audio video and text
  information,'' in \emph{Proceedings of the 7th Annual Workshop on
  Audio/Visual Emotion Challenge}, ser. AVEC ’17.\hskip 1em plus 0.5em minus
  0.4em\relax New York, NY, USA: Association for Computing Machinery, 2017, p.
  45–51. [Online]. Available: \url{https://doi.org/10.1145/3133944.3133950}
\BIBentrySTDinterwordspacing

\bibitem{Yin2019}
\BIBentryALTinterwordspacing
S.~Yin, C.~Liang, H.~Ding, and S.~Wang, ``A multi-modal hierarchical recurrent
  neural network for depression detection,'' in \emph{Proceedings of the 9th
  International on Audio/Visual Emotion Challenge and Workshop}, ser. AVEC
  ’19.\hskip 1em plus 0.5em minus 0.4em\relax New York, NY, USA: Association
  for Computing Machinery, 2019, p. 65–71. [Online]. Available:
  \url{https://doi.org/10.1145/3347320.3357696}
\BIBentrySTDinterwordspacing

\bibitem{Ray2019}
\BIBentryALTinterwordspacing
A.~Ray, S.~Kumar, R.~Reddy, P.~Mukherjee, and R.~Garg, ``Multi-level attention
  network using text, audio and video for depression prediction,'' in
  \emph{Proceedings of the 9th International on Audio/Visual Emotion Challenge
  and Workshop}, ser. AVEC ’19.\hskip 1em plus 0.5em minus 0.4em\relax New
  York, NY, USA: Association for Computing Machinery, 2019, p. 81–88.
  [Online]. Available: \url{https://doi.org/10.1145/3347320.3357697}
\BIBentrySTDinterwordspacing

\bibitem{Huang2020IS}
\BIBentryALTinterwordspacing
Z.~Huang, J.~Epps, D.~Joachim, B.~Stasak, J.~R. Williamson, and T.~F. Quatieri,
  ``{Domain Adaptation for Enhancing Speech-Based Depression Detection in
  Natural Environmental Conditions Using Dilated CNNs},'' in \emph{Proc.
  Interspeech 2020}, 2020, pp. 4561--4565. [Online]. Available:
  \url{http://dx.doi.org/10.21437/Interspeech.2020-3135}
\BIBentrySTDinterwordspacing

\bibitem{Browman1992}
C.~P. Browman and L.~Goldstein, ``{Articulatory Phonology : An Overview *},''
  \emph{Phonetica}, vol.~49, pp. 155--180, 1992.

\bibitem{Sivaraman2019}
\BIBentryALTinterwordspacing
G.~Sivaraman, V.~Mitra, H.~Nam, M.~Tiede, and C.~Espy-Wilson, ``Unsupervised
  speaker adaptation for speaker independent acoustic to articulatory speech
  inversion,'' \emph{The Journal of the Acoustical Society of America}, vol.
  146, no.~1, pp. 316--329, 2019. [Online]. Available:
  \url{https://doi.org/10.1121/1.5116130}
\BIBentrySTDinterwordspacing

\bibitem{Westbury1994a}
J.~R. Westbury, ``{Speech Production Database User ' S Handbook},'' \emph{IEEE
  Personal Communications - IEEE Pers. Commun.}, vol.~0, no. June, 1994.

\bibitem{Deshmukh2005}
O.~{Deshmukh}, C.~Y. {Espy-Wilson}, A.~{Salomon}, and J.~{Singh}, ``Use of
  temporal information: detection of periodicity, aperiodicity, and pitch in
  speech,'' \emph{IEEE Transactions on Speech and Audio Processing}, vol.~13,
  no.~5, pp. 776--786, 9 2005.

\bibitem{MUNDT2007}
\BIBentryALTinterwordspacing
J.~C. Mundt, P.~J. Snyder, M.~S. Cannizzaro, K.~Chappie, and D.~S. Geralts,
  ``Voice acoustic measures of depression severity and treatment response
  collected via interactive voice response (ivr) technology,'' \emph{Journal of
  Neurolinguistics}, vol.~20, no.~1, pp. 50 -- 64, 2007. [Online]. Available:
  \url{http://www.sciencedirect.com/science/article/pii/S0911604406000303}
\BIBentrySTDinterwordspacing

\bibitem{MUNDT2012}
\BIBentryALTinterwordspacing
J.~C. Mundt, A.~P. Vogel, D.~E. Feltner, and W.~R. Lenderking, ``Vocal acoustic
  biomarkers of depression severity and treatment response,'' \emph{Biological
  Psychiatry}, vol.~72, no.~7, pp. 580 -- 587, 2012, novel Pharmacotherapies
  for Depression. [Online]. Available:
  \url{http://www.sciencedirect.com/science/article/pii/S0006322312002636}
\BIBentrySTDinterwordspacing

\end{thebibliography}

\end{document}